\documentclass[aps,english,prl,amsmath,amsfonts,amssymb,superscriptaddress,twocolumn,showpacs,citeautoscript]{revtex4-1}
\usepackage[T1]{fontenc} \usepackage[latin9]{inputenc}
 \usepackage{amsmath} \usepackage{amssymb}
\usepackage{wasysym} \usepackage{esint} \usepackage{graphicx}
\usepackage{times} \usepackage{nicefrac} \usepackage{appendix}
\usepackage{textcase} \usepackage{subfigure} \usepackage{empheq}
\usepackage{color} \usepackage{leftidx} \usepackage{makecell}
\usepackage{stackrel} \makeatletter
\@ifundefined{textcolor}{}
{%
 \definecolor{BLACK}{gray}{0}
 \definecolor{WHITE}{gray}{1}
 \definecolor{RED}{rgb}{1,0,0}
 \definecolor{GREEN}{rgb}{0,1,0}
 \definecolor{BLUE}{rgb}{0,0,1}
 \definecolor{CYAN}{cmyk}{1,0,0,0}
 \definecolor{MAGENTA}{cmyk}{0,1,0,0}
 \definecolor{YELLOW}{cmyk}{0,0,1,0}
}

\renewcommand{\vec}[1]{\mathbf{#1}}
\renewcommand{\Re}{\operatorname{Re}}
\renewcommand{\Im}{\operatorname{Im}}

\newcommand{\sym}{\operatorname{sym}}

\newcommand{\citeasnoun}[1]{Ref.~\onlinecite{#1}}

\renewcommand{\eqref}[1]{(\ref{eq:#1})}

\newcommand{\Eqref}[1]{Equation~\ref{eq:#1}}
\newcommand{\figref}[1]{Fig.~\ref{fig:#1}}
\newcommand{\Figref}[1]{Figure~\ref{fig:#1}}

\def\ro{\vec{x}}
\def\rs{\vec{y}}
\newcommand{\mat}[1]{{#1}}

\newcommand{\Tr}{\text{Tr }}

\makeatother

\usepackage{babel}
\begin{document}
\title{General formulation of coupled radiative and conductive heat
  transfer between compact bodies}

\author{Weiliang Jin}
\affiliation{Department of Electrical Engineering, Princeton University, Princeton, NJ 08544, USA}
\author{Riccardo Messina}
\affiliation{Laboratoire Charles Coulomb, Universit\'{e} de Montpellier and CNRS, Montpellier, France}
\author{Alejandro W. Rodriguez}
\affiliation{Department of Electrical Engineering, Princeton University, Princeton, NJ 08544, USA}

\begin{abstract}
  We present a general framework for studying strongly coupled
  radiative and conductive heat transfer between arbitrarily shaped
  bodies separated by sub-wavelength distances. Our formulation is
  based on a macroscopic approach that couples our recent fluctuating
  volume--current (FVC) method of near-field heat transfer to the more
  well known Fourier conduction transport equation. We apply our
  technique to consider heat exchange between aluminum-zinc oxide
  nanorods and show that the presence of bulk plasmon resonances can
  result in extremely large radiative heat transfer rates (roughly
  twenty times larger than observed in planar geometries), whose
  interplay with conductive transport leads to nonlinear temperature
  profiles along the nanorods.
\end{abstract}
\maketitle

Radiative heat transfer (RHT) between objects held at different
temperatures can be many orders of magnitude larger in the near field
(short separations $d \ll$ thermal wavelength
$\lambda_T =\hbar c/k_\mathrm{B} T$) than for far-away
objects~\cite{basu2009review,volokitin2007near,ottens2011near,ReidRo12:review,song2015near}. Recently,
we showed that that the interplay of near-field RHT and conduction in
planar geometries can dramatically modify the temperature and thermal
exchange rate at sub-micron
separations~\cite{riccardo2016strongly}. Such strongly-coupled
conduction--radiation (CR) phenomena are bound to play a larger role
in situations involving structured materials, where RHT can be further
enhanced~\cite{miller2015shape,liu2015theory,phan2013near,yang2016spectrally,biehs2012hyperbolic,dai2015enhanced}
and
modified~\cite{rodriguez2013anomalous,rodriguez2012fluctuating,sheila2016surface,rodriguez2011frequency},
and in on-going experiments exploring nanometer scale gaps, where the
boundary between conductive (phonon- and electron-mediated) and
radiative transport begins to
blurr~\cite{cahill2014nanoscale,chiloyan2015natcomm}.

We present a general CR framework that captures the interplay of
near-field RHT and thermal conduction along with the existence of
large temperature gradients in arbitrary geometries. We show that
under certain conditions, i.e. materials and structures with
separations and geometric lengthscales in the nanometer range, RHT can
approach and even exceed conduction, significantly changing the
stationary temperature distribution of heated objects. Our approach is
based on a generalization of our recent fluctuating volume-current
(FVC) formulation of electromagnetic (EM) fluctuations, which when
coupled to the more standard Fourier heat equation describing
conductive transport at macroscopic scales, allows studies of CR
between arbitrary shapes, thereby generalizing our prior work with
slabs~\cite{riccardo2016strongly}. As a proof of concept, we consider
an example geometry involving aluminum-zinc oxide (AZO) nanorods
separated by vacuum gaps, which exhibits more than an order of
magnitude enhancement in RHT compared to planar slabs, and hence leads
to even larger temperature gradients. We find that while RHT between
thin slabs is primarily mediated by surface modes, resulting in linear
temperature gradients, the presence of bulk nanorod resonances leads
to highly distance-dependent nonlinear temperature profiles.

Coupled radiative and conductive diffusion processes in nanostructures
are becoming increasingly
important~\cite{joulain2008near,chiloyan2015natcomm}, with recent
works primarily focusing on the interplay between thermal diffusion
and external optical illumination such as laser-heating of plasmonic
structures~\cite{baffou2010mapping,baffou2014deterministic,ma2014heat,baldwin2014thermal,biswas2015sudden}. On
the other hand, while it is known that conduction has a strong
influence on RHT experiments~\cite{st2016near,st2014demonstration},
the converse has thus far been largely unexplored because RHT is
typically too small to result in appreciable temperature
gradients~\cite{wong2011monte,wong2014coupling,lau2016parametric}. However,
our recent work~\cite{riccardo2016strongly} suggests that such an
interplay can be significant at tens of nanometer separations and in
fact may already have been present (though overlooked) in recent
experiments involving planar
systems~\cite{shen2009surface,kittel2005near,kim2015radiative,song2015enhancement}. Moreover,
since planar structures are known to exhibit highly suboptimal RHT
rates~\cite{miller2015shape}, we expect even stronger interplays in
more complex geometries, such as metasurfaces~\cite{liu2015near},
hyperbolic metamaterials~\cite{liu2014nearmeta,biehs2012hyperbolic},
or lattices of metallic antennas~\cite{liu2015theory,miller2015shape}.

\begin{figure*}[t!]
\begin{center}
\includegraphics[width=1.95\columnwidth]{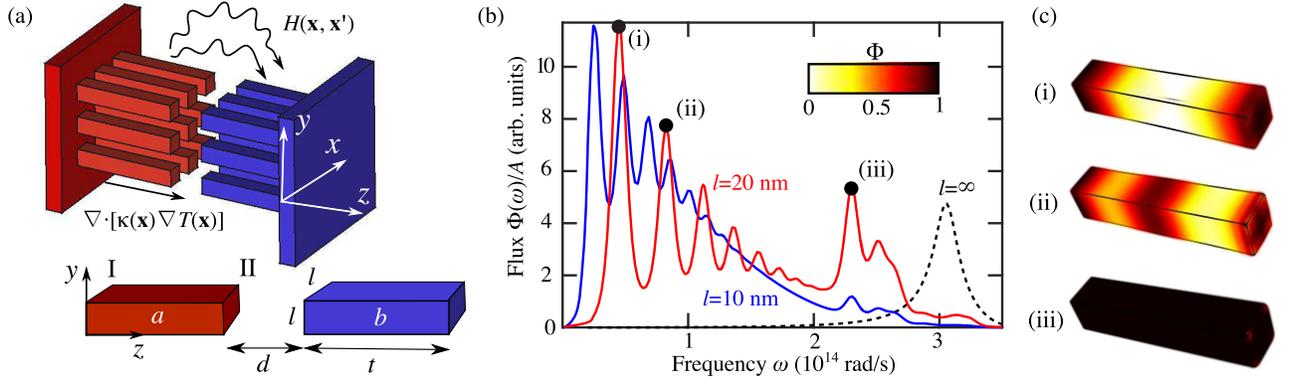}
\caption{(a) Schematic illustration of two square lattices of nanorods
  (labelled $a$ and $b$) of thickess $t$, period $\Lambda$,
  cross-sectional area $l\times l$, and separation $d$, whose
  temperature distribution and energy exchange is mediated by both
  conductive $\nabla\cdot[\kappa(\mathbf{x})T(\mathbf{x})]$ and
  radiative $H(\mathbf{x},\mathbf{x}')$ heat transfer. (b) Total
  radiative heat transfer spectrum $\Phi(\omega)$ between two AZO
  nanorods (solid lines) of thickness $t=500$~nm and cross-sectional
  area $A=l^2$, separated by $d=20$~nm and held at temperatures
  $T_{a(b)}=800(300)$~K. The spectrum is shown for different
  cross-sections $l=\{10,20\}$~nm (blue and red lines) and in the
  limit $l = \infty$, corresponding to two planar slabs. (c) Spatial
  radiative heat flux in nanorod $a$ for the case $l=20$~nm,
  corresponding to the (i) first, (ii) second, and (iii) SPP plasmon
  resonances, respectively, annotated in (b).}
\label{fig:fig1}
\end{center}
\end{figure*}

\emph{Formulation.---} In what follows, we describe a general
formulation of coupled CR applicable to arbitrary geometries. Consider
a situation involving two bodies (the same framework can be extended
to multiple bodies), labelled $a$ and $b$, subject to arbitrary
temperature profiles and exchanging heat among one other, shown
schematically in \figref{fig1}(a). Neglecting convection and
considering bodies with lengthscales larger or of the order of their
phonon mean-free path, in which case Fourier conduction is valid, the
stationary temperature distribution satisfies:
\begin{equation}
  \nabla\cdot[\kappa(\vec{x})\nabla T(\vec{x})]+
  \int\mathrm{d}^3\vec{x}'\,H(\vec{x},\vec{x}')=Q(\vec{x})
\label{eq:heat}
\end{equation}
where $\kappa(\vec{x})$ and $Q(\vec{x})$ describe the bulk Fourier
conductivity and presence of some external heat source at $\vec{x}$,
respectively, and $H(\vec{x},\vec{x}')$ denotes the radiative power
per unit volume from $\vec{x}'$ to $\vec{x}$.

Our ability to compute $H(\vec{x},\vec{x}')$ in full generality hinges
on an extension of a recently introduced FVC method that exploits
powerful EM scattering techniques~\cite{polimeridis2015fluctuating} to
enable fast calculations of RHT under arbitrary geometries and
temperature distributions. The starting point of this method is the
volume-integral equation (VIE) formulation of EM, in which the
scattering unknowns are 6-component polarization currents $\xi$ in the
interior of the bodies coupled via the homogeneous $6\times 6$ Green's
function $\Gamma$ of the intervening
medium~\cite{polimeridis2015fluctuating}. Given two objects described
by a susceptibility tensor $\chi(\vec{x})$ and a Galerkin
decomposition of the induced currents $\xi=\sum_i x_i b_i$, with
$\{b_i\}$ denoting localized basis functions throughout the objects
($i$ is the global index for all bodies), the scattering of an
incident field due to some fluctuating current-source $\sigma=\sum_i
s_i b_i$ can be determined via solution of a VIE equation, $x+s=Ws$,
in terms of the unknown and known expansion coefficients $\{x_i\}$ and
$\{s_i\}$, respectively, where $W^{-1}_{i,j}=\langle b_i,
(I+i\omega\chi G) b_j\rangle$ and $G_{i,j}=\langle b_{i},\Gamma\star
b_{j}\rangle$ are known as VIE and Green
matrices~\cite{polimeridis2015fluctuating}.  Previously, we exploited
this formalism to propose an efficient method for computing the total
heat transfer between any two compact
bodies~\cite{polimeridis2015fluctuating}, based on a simple
voxel basis expansion (uniform discretization).  The solution of
\eqref{heat} requires an extension of the FVC method to include the
spatially resolved heat transfer between any two voxels, which we
describe below.

Consider a fluctuating current-source
$\sigma_{\alpha}=s_{\alpha}b_{\alpha}$ at $\mathbf{x}_a=b_{\alpha}$ in
a body $a$. Such a ``dipole'' source induces polarization--currents
$\xi_{\beta}=x_{\beta}b_{\beta}$ and EM fields $\phi_{\beta}$
throughout space in body $b$ (and elsewhere), such that the heat flux
at $\mathbf{x}_b=b_{\beta}$ is given (by Poynting's theorem) by:
\begin{align}
  \Phi(\omega;\mathbf{x}_a\rightarrow\mathbf{x}_b)=\frac{1}{2}\langle\Re \left( \xi_{\beta}^{*}\phi_{\beta} \right)\rangle
\label{eq:Phi1}
\end{align}
where ``$\langle \ldots \rangle$'' denotes a thermodynamic ensemble
average.  Expressing the polarization--currents and fields in the
localized basis $\{b_{\alpha}\}$, and exploiting the volume
equivalence principle to express the field as a convolution of the
incident and induced currents with the vacuum Green's function (GF),
$\phi=\Gamma\star (\xi+\sigma)$, one finds that \eqref{Phi1} can be
expressed in a compact, algebraic form involving VIE matrices:
\begin{align}
  \Phi(\omega;\mathbf{x}_a\rightarrow\mathbf{x}_b)&=\frac{1}{2}\langle\Re \left\{  
    x_{\beta}^{*}[G(x+s^\alpha)]_{\beta}\right\}\rangle\nonumber\\
&=\frac{1}{2} \langle\Re \left\{  (x+s^\alpha)_{\beta}^{*}[G(x+s^\alpha)]_{\beta}\right\}\rangle\nonumber\\
&=\frac{1}{2}\langle \Re \left[  (Ws^\alpha)_{\beta}^{*}(GWs^\alpha)_{\beta}\right]\rangle\nonumber\\
&=\frac{1}{2} \Re \left[  D_{\alpha,\alpha}W^{\dagger}_{\alpha,\beta}(GW)_{\beta,\alpha}\right]
\label{eq:Phi3}
\end{align}
where $s^{\alpha}$ is a vector that is zero everywhere except at the
$\alpha$th element, denoted by $s_{\alpha}$, and
$\mat{D}_{\alpha,\beta}=\langle s_{\alpha}^{*}s_{\beta}\rangle=\int
\int d^3\ro\,d^3\rs \, b_{\alpha}^\ast(\ro) \langle \sigma(\ro)
\sigma^\ast(\rs) \rangle b_{\beta}(\rs)$
is a real, diagonal matrix encoding the thermodynamic and dissipative
properties of each object~\cite{polimeridis2015fluctuating} and
described by the well-known fluctuation--dissipation theorem,
$\langle \sigma_i(\ro,\omega) \sigma_j^\ast(\rs,\omega) \rangle =
\frac{4}{\pi} \omega\Im \epsilon(\ro,\omega) \Theta (T_{\vec{x}})
\delta(\ro-\rs)\delta_{ij}$,
where $\Theta(T)=\hbar\omega/[\mathrm{exp}(\hbar\omega/k_bT)-1]$ is
the Planck distribution. It follows then that the heat flux emitted or
absorbed at a given position $\mathbf{x}_a$, the main quantity
entering \eqref{heat} through
$\int\mathrm{d}^3\vec{x}'\,H(\vec{x},\vec{x}')=\int \mathrm{d}\omega\,
\Phi(\omega;\mathbf{x})$, is given by:
\begin{align}
  \Phi(\omega;\mathbf{x}_a)&=\int_{V_{b}} \mathrm{d}^3\mathbf{x}_b\left[ \Phi(\omega;\mathbf{x}_b\rightarrow
\mathbf{x}_a)-\Phi(\omega;\mathbf{x}_a\rightarrow
\mathbf{x}_b)\right]\nonumber\\
&=\frac{1}{2}\Tr_{\beta | b_\beta \in V_b} \Re \left[
  D_{\beta,\beta}W^{\dagger}_{\beta,\alpha}(GW)_{\alpha,\beta}-(\alpha\leftrightarrow\beta)\right]\nonumber\\
  &=\frac{1}{2}\Re \left[\underbrace{GWD^bW^{\dagger}}_{\Phi_\mathrm{a}}-\underbrace{DW^{\dagger}P^bGW}_{\Phi_\mathrm{e}}\right]_{\alpha,\alpha}
\label{eq:Phi2}
\end{align}
Here, $P^{a(b)}$ denotes the projection operator that selects only
basis functions in $a(b)$, such that $D^b=P^bDP^b$ is a diagonal
matrix involving only fluctuations in object $b$.  Furthermore, the
first (second) term in \eqref{Phi2} describe the absorbed (emitted)
power in $\vec{x}_a$, henceforth denoted via the subscript ``a(e)''.

\begin{figure*}[t!]
\begin{center}
\includegraphics[width=2\columnwidth]{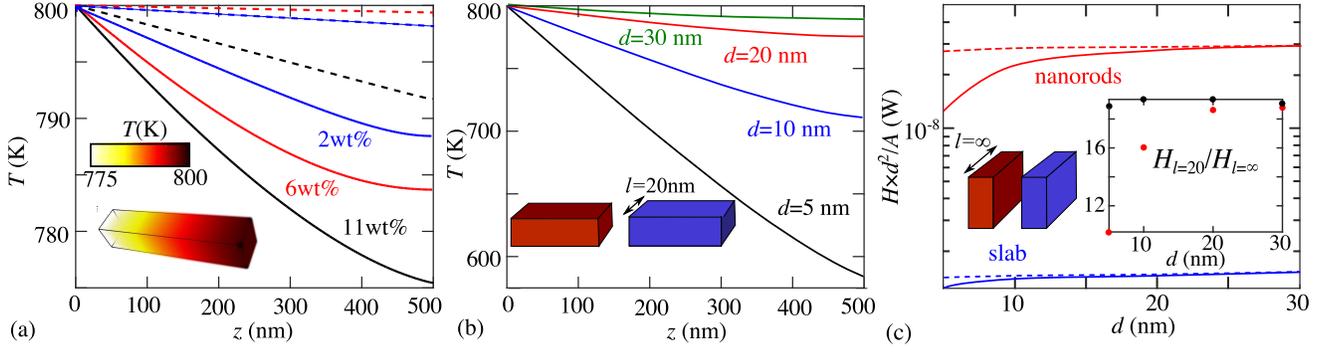}
\caption{(a) Temperature profile along the $z$ coordinate of a nanorod
  (solid lines) when it is heated from one side to a temperature of
  800~K, and is separated from an identical, constant- and
  uniform-temperature nanrod held at $T=300$~K on the other side, by a
  gap size $d=20$~nm. The nanorods have cross-sectional width
  $l=10$~nm and thicknesses $t=500$~nm, and are made up of AZO with
  results shown for multiple values of the doping concentration
  $\{2,6,11\}\mathrm{wt}\%$ (blue, red, and black lines). Also shown
  are the temperature profiles of slabs (dashed lines) of the same
  thickness (corresponding to the limit $l \to \infty$).  (Inset:)
  Temperature distribution throughout the nanorod in the case of
  $11\mathrm{wt}\%$. (b) Temperature profiles of nanorods of width
  $l=20$~nm under various separations $d=\{5,10,20,30\}$~nm (black,
  blue, red, and green lines). (c.inset:) The ratio of total radiative
  heat flux for nanorods of width $l=20$~nm to that of the slabs as a
  function of $d$, in the presence (red dots) or absence (black dots)
  of temperature gradients induced by conduction and radiation
  interplay, with the flux value shown in (c), for nanorods (red) and
  slabs (blue), also in the presence (solid lines) or absence (dashed
  lines) of temperature gradients induced by the interplay of
  conduction and radiation.}
\label{fig:result}
\end{center}
\end{figure*}

\Eqref{Phi2} is a generalization of our previous expression for the
total heat transfer between two arbitrary inhomogeneous
objects~\cite{polimeridis2015fluctuating} in that it includes both the
spatially resolved absorbed and emitted power throughout the entire
geometry. In~\citeasnoun{polimeridis2015fluctuating}, we showed that
the low-rank nature of the GF operator enables truncated, randomized
SVD factorizations and therefore efficient evaluations of the
corresponding matrix operations. We find, however, that in this case,
the inclusion of the absorption term does not permit such a
factorization, except in special circumstances. In particular, writing
down the two terms separately by expanding into the subspace spanned
by each object, we find:
\begin{align}
\label{eq:Phia}
  \Phi_\mathrm{a}(\omega;\mathbf{x}_a)&=\frac{1}{2}\Re \left[
    G^{ab}W^{bb}D^{bb}W^{ab\dagger}+G^{aa}W^{ab}D^{bb}W^{ab\dagger}\right] \\
  \Phi_\mathrm{e}(\omega,\vec{x}_a) &= -\frac{1}{2}\left[ \Re(D^{aa}
    W^{ba\dagger} G^{ba} W^{aa})\right. \nonumber \\
    &\hspace{0.7in}\left.+ D^{aa} W^{ba\dagger} \sym(G^{bb}) W^{ba} \right]_{\alpha,\alpha} 
\label{eq:Phie}
\end{align}
with $X^{ij} = P^i X P^j$ denoting the sub-block of matrix $X$
connecting basis functions in object $i$ to object $j$, and $\sym X =
\frac{1}{2}(X+X^\dagger)$ denoting the symmetric part of $X$.

\Eqref{Phie}, describing emission, can be evaluated efficiently
because the matrices $G^{ba}$ and $\sym{G^{bb}}$ are both low rank
($\ell \ll N$)~\cite{polimeridis2015fluctuating}, in which case they
can be SVD factorized to allow fast matrix multiplications. It follows
that the total heat transfer, i.e. the trace of \eqref{Phie}, can also
be computed efficiently. Unfortunately, the second term of
\eqref{Phia} involves both the symmetric and anti-symmetric parts of
$G^{aa}$, the latter of which is full rank. More conveniently,
detailed balance dictates that $\Phi(\omega;\mathbf{x}_b\rightarrow
\mathbf{x}_a)=\Phi(\omega;\mathbf{x}_a\rightarrow \mathbf{x}_b)$
whenever $T(\mathbf{x}_a) = T(\mathbf{x}_b)$, which implies that $\Re
\left[ M_{\beta,\beta}
  W^{\dagger}_{\beta,\alpha}(GW)_{\alpha,\beta}\right]=\Re
\left[M_{\alpha,\alpha}
  W^{\dagger}_{\alpha,\beta}(GW)_{\beta,\alpha}\right]$, where
$M_{\alpha,\alpha}=\Im \varepsilon(\mathbf{x}_{\alpha},\omega)$ is a
real, diagonal matrix encoding the dissipative properties of the
bodies, leading to the following modified expression for the
absorption rate:
\begin{align}
  \Phi_\mathrm{a}(\omega;\mathbf{x}_a)&=\frac{1}{2}\left[\Re(M^{aa}W^{ba\dagger}K^{bb}G^{ba}W^{aa})\right.\nonumber\\
  &\hspace{0.7in}\left.+M^{aa}W^{ba\dagger}\sym(K^{bb}G^{bb})W^{ba}\right]_{\alpha,\alpha}
  \label{eq:absorb}
\end{align}
where the real and diagonal matrix
$K_{\alpha,\alpha}=D_{\alpha,\alpha}/M_{\alpha,\alpha}$ is only
relevant to the Plank function
$\Theta(T(\mathbf{x}_{\alpha}),\omega)$. Noticeably, the symmetrized
operator in the second term is full rank except whenever the
temperature of object $b$ is close to uniform, in which case $\sym
(K^{bb} G^{bb}) \approx K^{bb} \sym G^{bb}$. While solution of
\eqref{absorb} is feasible, it remains an open problem to find a
formulation that allows fast evaluations of the spatially resolved
absorbed power under arbitrary temperature distributions.

Given \eqref{Phi2}, one can solve the coupled CR equation in any
number of ways~\cite{press2007numerical}. Here, we exploit a
fixed-point iteration procedure based on repeated and independent
evaluations of \eqref{Phi3} and \eqref{heat}, converging once both
quantities approach a set of self-consistent steady-state
values. \Eqref{heat} is solved via a commercial, finite-element heat
solver whereas \eqref{Phi3} is solved through a free, in-house
implementation of our FVC method~\cite{polimeridis2015fluctuating}.
While the above formulation is general, below we explore the
computationally convenient situation in which object $b$ is kept at a
constant, uniform temperature by means of a carefully chosen thermal
reservoir, such that the absorbed power in object $a$ can be computed
efficiently via~\eqref{absorb}. Furthermore, absorption can be
altogether ignored whenever one of the bodies is heated to a much
larger temperature than the other (as is the case below). The power
emitted by $a$ (the heated object), obtained via \eqref{Phie}, turns
out to be much more convenient to compute, since the time-consuming
part of the scattering calculation can be precomputed independently
from the temperature distribution and stored for repeated and
subsequent evaluations of \eqref{heat} under different temperature
profiles.

\emph{Results.---} As a proof of principle and to gain insights into
coupled CR effects in non-planar objects, we now apply the above
method to a simple geometry consisting of two metallic nanorods of
cross-sectional widths $l$ and thickness $t$; in practice, both for
easy of fabrication and to obtain even larger RHT~\cite{st2016near},
such a structure could be realized as a lattice or grating, shown
schematically in \figref{fig1}(a). However, for computational
convenience and conceptual simplicity, we restrict our analysis to the
regime of large grating periods, in which case it suffices to consider
only the transfer between nearby objects.  The strongest CR effects
generally will arise in materials that exhibit large RHT,
e.g. supporting surface--plasmon polaritons (SPP) in the case of
planar objects, and low thermal conductivities, including silica,
sapphire, and AZO, whose typical thermal conductivities
$\sim 1$~W/m$\cdot$K. In the following, we take AZO as an illustrative
example~\cite{loureiro2014transparent,naik2011oxides}. To begin with,
we show that even in the absence of CR interplay, the RHT spectrum and
spatial RHT distribution inside the nanorods differ significantly from
those of AZO slabs of the same thickness.

\Figref{fig1}(b) shows the RHT spectrum $\Phi(\omega)$ per unit area
$A=l^2$ between two AZO nanorods (with doping concentration
$11\mathrm{wt}\%$~\cite{naik2011oxides}) of length $t=500$~nm and
varying widths $l=\{10,20,\infty\}$~nm (blue solid, red solid, and
black dashed lines), held at temperatures $T_{a(b)}=800(300)$~K and
vacuum gap $d=20$~nm. The limit $l\to\infty$ corresponds to the
slab-slab geometry already explored~\cite{riccardo2016strongly}, in
which case the $\Phi(\omega)$ exhibits a single peak occuring at the
SPP frequency $\approx 3\times 10^{14}$~rad/s. The finite nature of
the nanorods results in additional peaks at lower frequencies,
corresponding to bulk/geometric plasmon resonances (red and blue solid
lines) that provide additional channels of heat exchange, albeit at
the expense of weaker SPP peaks, leading to a roughly $20$-fold
enhancement in RHT compared to slabs.  More importantly and well
known, such structured antennas allow tuning and creation of bulk
plasmon resonances in the near- and far-infrared spectra (much lower
than many planar materials) that can more effectively transfer thermal
radiation. The contour plots in \figref{fig1}(i--iii) reveal the
spatial RHT distribution $\Phi(\omega,\mathbf{x})$ (in arbitrary
units) at three separate frequencies
$\omega=\{0.4,0.8,2.3\}\times 10^{14}$~rad/s, corresponding to the
first, second, and SPP resonances, respectively. As expected, the
highest-frequency resonance is primarily confined to the corners of
the nanorod surface (becoming the well-known SPP resonance in the
limit $l \to \infty$), with the fundamental and intermediate
resonances have flux contributions stemming primarily form the
bulk. As we now show, such an enhancement results not only results in
larger temperature gradients but also changes the resulting
qualitative temperature distribution.

\Figref{result}(a) shows the temperature profile along the $z$
direction for the nanorod geometry of \figref{fig1}(a), with width
$l=10~$nm and gap size $d=20$~nm, obtained via solution of
\eqref{heat}. For the purpose of generality, we show results under
various doping concentrations $\{2,6,11\}\mathrm{wt}\%$ (green, red,
and black solid lines), corresponding to different SPP frequencies and
bandwidths~\cite{naik2011oxides}. In particular, we consider a
situation in which the boundary I of nanorod $a$ is kept at
$T_\text{I}=800$~K while the entire nanorod $b$ is held at $T_b=300$~K
(through contact with a room-temperature reservoir), and assume an AZO
thermal conductivity of
$\kappa=1$~W/m$\cdot$K~\cite{loureiro2014transparent}. The temperature
along the $x$--$y$ cross section is nearly uniform (due to the faster
heat diffusion rate along the smaller dimension) and therefore only
shown in the case of $11\mathrm{wt}\%$ (inset). In all scenarios, the
temperature gradient is significantly larger for nanorods (solid
lines) than for slabs ($t \to \infty$, dashed lines), becoming an
order of magnitude larger in the case of $6\mathrm{wt}\%$ due to its
larger SPP frequency compared to the peak Planck wavelength near
$800$~K. Furthermore, while slabs exhibit linear temperature profiles
(RHT is dominated by surface emission~\cite{basu2009review}), the
bulky and de-localized nature of emission in the case of nanorods
results in nonlinear temperature distributions.

\Figref{result}(b) shows the temperature profile at various
separations $d=\{5,10,20,30\}$~nm (black, blue, red, and green lines)
for nanorods of width $l=20$~nm and $11\mathrm{wt}\%$, illustrating
the sensitive relationship between the degree of CR interplay and gap
size. Notably, while the RHT and therefore temperature gradients
increase as $d$ decreases, the profile becomes increasingly linear as
the geometry approaches the slab--slab configuration. The transition
from bulk- to surface-dominated RHT and the increasing impact of the
latter on conduction and vice versa is also evident from
\figref{result}(c). The figure shows the radiative flux rate $H\times
d^2$ as a function of $d$ for slabs (black lines) of thickness
$t=500$~nm and nanorods (red lines) of equal thickness and width
$l=20$~nm, either including (solid lines) or excluding (dashed lines)
CR interplay (with the latter involving uniform temperatures). While
the RHT between bodies of uniform temperatures is shown to scales as
$1/d^2$ (dashed lines), the temperature gradients induced by CR
interplay in the case of nanorods begins to change the expected
powerlaw behavior at $d\approx 15$~nm; the same occurs for slabs but
at much shorter $d\lesssim 5$~nm. These differences are further
quantified on the inset of the figure, which shows the ratio of the
RHT rate between the two objects as a function of $d$. While the ratio
remains almost a constant for uniform-temperature objects (black
dots), it decreases visibly when considering CR interplay (red
dots). As shown in~\citeasnoun{riccardo2016strongly}, in the limit
$d\rightarrow 0$, RHT will asymptote to a constant (not shown) rather
than a diverge.

\emph{Concluding remarks.---} As experiments continue to push toward
larger RHT by going to smaller vacuum gaps or through nanostructuring,
accurate descriptions of CR interplay and associated effects will
become increasingly
important~\cite{chiloyan2015natcomm,cahill2014nanoscale}. Future work
along these directions could focus on extending our work to periodic
structures, which could potentially exhibit much larger RHT and hence
CR effects.

\emph{Acknowledgements.---} This work was supported by the National
Science Foundation under Grant no. DMR-1454836 and by the Princeton
Center for Complex Materials, a MRSEC supported by NSF Grant DMR
1420541.

\end{document}